\documentclass[10pt]{article}

\usepackage{graphicx}
\usepackage{indentfirst,csquotes}
\usepackage{amssymb,amsthm,amsmath}
\usepackage{xcolor,paralist,hyperref,titlesec,fancyhdr}
\usepackage{lipsum}
\usepackage{siunitx}    
\usepackage{booktabs}   
\usepackage{float}

\titleformat{\section}{\normalfont\large\bfseries}{\thesection}{1em}{}
\titlespacing*{\section}{0pt}{1ex}{1ex}

\hypersetup{ colorlinks=true, linkcolor=black, filecolor=black, urlcolor=black }

\begin{document}

\title{Observation of the Aharonov--Bohm Effect in Pilot-Wave Hydrodynamics}

\author{
  Georgi Gary Rozenman$^{1}$ \and
  Kyle I. McKee$^{1}$ \and
  Arnaud Lazarus$^{1}$ \and
  Valeri Frumkin$^{2}$ \and
  John W. M. Bush$^{1}$
}

\date{%
  $^{1}$ Department of Mathematics, Massachusetts Institute of Technology, 
  Cambridge, Massachusetts, USA\\
  $^{2}$ Department of Mechanical Engineering, Boston University, 
  Boston, Massachusetts, USA\\[1ex]
  \texttt{garyrozenman@protonmail.com}\\[2ex]
  \today
}

\maketitle

\begin{abstract}
\quad
We report the results of an experimental study of an analog of the Aharonov-Bohm (AB) effect achieved with the hydrodynamic pilot-wave system.  A walking droplet is confined to an annular cavity that encircles a shielded vortex, but lies outside its range of direct influence. While there is no vortex-induced flow in the immediate vicinity of the droplets, the vortex modifies the droplet's spatially extended pilot-wave field that guides its motion, producing a vortex-dependent bias in the droplet’s orbital speed. High-speed tracking and delay-embedding reconstructions yield Wigner-like phase-space distributions for this hydrodynamic system that exhibits a rigid, flux-dependent translation, providing a force-free, gauge-like realization of an AB-type phase.
\end{abstract} 

\bigskip

\paragraph*{Introduction.}

In classical mechanics, a particle’s state is specified by its instantaneous position and momentum, whose evolution is governed by local forces. 
Quantum mechanics provides a fundamentally different description that informs only the statistical 
behavior of microscopic systems \cite{feynman1965feynman}. 
Observable outcomes arise from the coherent superposition of amplitudes associated with all possible paths, so that differences in the accumulated action manifest as measurable phase shifts.
In 1959, Aharonov and Bohm demonstrated that a charged particle can acquire a relative phase even when it propagates only through regions where the magnetic field vanishes \cite{Aharonov1959}. 

In the canonical two-slit geometry emphasized in their original paper, an incident beam is split into two coherent partial waves that pass on opposite sides of an idealized, infinitely long solenoid placed between the slits and the screen. 
Although $\mathbf{B}=0$ along both paths, the vector potential $\mathbf{A}$ is nonzero in the accessible region, and the two partial waves accumulate different path integrals $\int \mathbf{A}\cdot d\boldsymbol{\ell}$ before recombining. 
The resulting phase difference depends only on the enclosed flux and on the topology of the paths, not on local forces, and is therefore geometric in character: it is an associated holonomy  with transport in the gauge potential. 
Consequently, the interference fringes shift by an amount set by the enclosed flux, providing an operational signature of a gauge potential through a purely path-dependent phase.
\begin{equation}
\varphi_{\mathrm{AB}} = \frac{\Delta S}{\hbar},
\end{equation}
where $\Delta S$ is the difference in the classical action $S=\int L\,dt$ accumulated along the two interfering paths and $\hbar$ is the reduced Planck constant. This phase shift arises even when the magnetic field vanishes along each path \cite{Aharonov1959}.
Because the electromagnetic vector potential enters the action through minimal coupling, it can imprint an observable phase on the wavefunction despite the absence of any local Lorentz force, the inference being that gauge potentials---not only fields---have physical consequences in quantum systems \cite{AharonovBohm1961}.

A wealth of experiments has confirmed the Aharonov–Bohm effect in microscopic systems \cite{Battalani}. Exmaples include Tonomura \textit{et al.}’s electron holography loops around a toroidal magnet \cite{TonomuraABeffect}, the $h/e$ and $h/2e$ conductance oscillations of mesoscopic Cu rings reported by Webb, Washburn, and co-workers \cite{WebbWashburn1985,WashburnWebbAdvPhys1986,MohantyDecoherence1997}, and more recent interference effects in semiconductor quantum rings \cite{FuhrerNature2001}, graphene corrals \cite{AokiPhysicaE2008,HuefnerPSSB2009}, and topological-insulator nanowires \cite{PengNatMater2010,BardarsonPRL2010}.
These studies unambiguously establish the observable consequences of the phase shift predicted by AB theory. However, complementary signatures of the AB-effect remain experimentally demanding. First, a direct visualization of the flux-induced displacement in momentum or phase space has yet to be reported. Second, a concurrent, real-time measurement of the kinetic-energy offset that accompanies the canonical-momentum shift in finite-size geometries is parametrically small for experimentally accessible fluxes. Third, the single-particle, single-shot determination of the enclosed flux proposed by Aharonov and co-authors \cite{EPL110_50004} remains experimentally challenging. We here adopt the hydrodynamic analog, which captures a platform for probing these features of the AB effect.

Following its original formulation, the Aharonov–Bohm effect has also been interpreted in terms of wavefront dislocations. Berry showed that the AB phase may be viewed as arising from a topological defect in the wavefront structure and proposed a direct analogy with water-wave systems \cite{berry1980wavefront}. This perspective emphasizes the geometric, path-dependent character of the effect and motivates the exploration of Aharonov-Bohm-like physics in other classical wave-mediated systems. We here employ the hydrodynamic pilot-wave system, a macroscopic realization of wave-particle duality in which a particle is accompanied by a spatially extended wave field, and so explore whether an \emph{analogous gauge phase} can arise in a macroscopic, classical driven system.

In pilot-wave hydrodynamics, a millimetric droplet bounces on a vertically vibrated bath of silicone oil~\cite{Couder2005,Bush2015}. In certain parameter regimes, they achieve resonance with their Faraday wave field and propel themselves along the bath surface, dressed in a quasi-monochromatic wave field. This system has produced an impressive catalog of quantum-like effects~\cite{Bush2020,bush2024}, including single-particle diffraction~\cite{Couder2006,Pucci2018}, tunneling~\cite{Eddi2011}, quantized orbits~\cite{FortEtAl2010,HarrisBush2014,Perrard2014}, Anderson localization~\cite{Abraham2024} and statistical projection effects in corrals~\cite{Saenz2018Statistical}. 
Owing to the identical forms of the Lorentz force that acts on a moving charge in a uniform magnetic field and the Coriolis force that acts on a mass in a rotating frame, vorticity induced by uniform rotation has been used as a proxy for magnetic field in a number of hydrodynamic quantum analogs, including Larmor levels ~\cite{FortEtAl2010} and spin lattices~\cite{Saenz2021SpinLattices}.
We here consider walkers interacting with a shielded vortex that plays the role of a magnetic solenoid in the AB effect. A key result of our study is the identification of a regime in which the vortex is dynamically isolated from the droplet: the walker’s trajectory lies entirely outside the range of influence of the vortex core, where the mean surface flow is negligible, so there is no direct hydrodynamic force acts on the particle. Nevertheless, the vortex modifies the pilot-wave field that mediates the droplet’s propulsion, leading to a systematic shift of the droplet’s momentum. 
In this regime, the droplet’s dynamics are altered solely through a geometric phase accumulated by its guiding wave, while the droplet itself experiences no local force from the vortex. This phase-only modification provides a direct classical analogue of a gauge potential in the sense of the Aharonov–Bohm effect. Despite the many quantum analogs realized with walking droplets, a clear demonstration of such a force-free, gauge-like influence — and its manifestation in phase space — has yet to be realized.

\textit{Theoretical background.}

For a charged particle of mass $m$ and charge $q$ constrained to a ring of radius $R$ threaded by a magnetic flux $\Phi$, the Aharonov-Bohm effect may be formulated most transparently in terms of the vector potential. In polar coordinates $(r,\theta)$, a convenient choice is the Coulomb gauge,
\begin{equation}
\mathbf{A}(\mathbf{r})=\frac{\Phi}{2\pi r}\,\hat{\boldsymbol{\theta}},
\quad\Rightarrow\quad
A_{r}=0,\;\;A_{\theta}=\frac{\Phi}{2\pi r},
\label{eq:CoulombGauge}
\end{equation}
for which the magnetic field vanishes everywhere along the particle trajectory. Nevertheless, a particle encircling the flux acquires a nontrivial phase
\begin{equation}
\Delta\varphi
=\frac{q}{\hbar}\oint_{\mathcal{C}}\mathbf{A}\cdot d\boldsymbol{\ell}
=\frac{q\Phi}{\hbar},
\label{eq:ABphase_action}
\end{equation}
demonstrating that the gauge potential can influence the wavefunction even in the absence of any local magnetic field along the path.

When the particle is confined to the ring $(s=R\theta)$, its dynamics are governed by the Hamiltonian
\begin{equation}
H=\frac{1}{2mR^{2}}\!\left(-i\hbar\frac{\partial}{\partial\theta}-\frac{q\Phi}{2\pi}\right)^{2},
\label{eq:ringH}
\end{equation}
whose eigenenergies are
\begin{equation}
E_{n}=\frac{\hbar^{2}}{2mR^{2}}\!\left(n-\frac{\Phi}{\Phi_{0}}\right)^{2},\qquad 
\Phi_{0}\equiv\frac{h}{q},
\label{eq:ringSpectrum}
\end{equation}
with $n\in\mathbb{Z}$ the angular-momentum (winding-number) quantum number. The flux dependence of the spectrum encodes the Aharonov--Bohm phase and underlies the characteristic shift of dynamical observables with enclosed flux~\cite{Aharonov1959}.

At the level of kinematics, the coupling between the particle and the enclosed flux 
produces a uniform shift of the canonical momentum,
\begin{equation}
\Delta p = qA_{\theta}=\frac{q\Phi}{2\pi R},
\label{eq:Deltap}
\end{equation}
despite the magnetic field being zero along the ring. This separation between a nonzero gauge potential and a vanishing local field is the defining feature of the Aharonov--Bohm effect. Specifically, while the particle trajectory lies outside the region of non-negligible field, the magnetic flux enclosed by the trajectory generates an effective gauge potential that shifts the canonical momentum.

The corresponding velocity expectation value follows directly from the quantum-mechanical velocity operator. For a single angular-momentum branch, the longitudinal velocity expectation value takes the form
\begin{equation}
\langle v_x(t)\rangle = v_0 \cos(\omega t),
\label{eq:vx_expectation}
\end{equation}
where the flux-dependent velocity amplitude
\begin{equation}
v_0=\frac{\hbar n}{mR}-\frac{q\Phi}{2\pi m R}
\label{eq:flux}
\end{equation}
reflects the shift of the canonical momentum in Eq.~\eqref{eq:Deltap}, and the angular frequency is given by $\omega=v_0/R$ (see, e.g., Ref.~\cite{Griffiths}). Equation~\eqref{eq:vx_expectation} provides a valuable point of comparison for the experiments presented below.


\paragraph*{Hydrodynamic analog.}
This ring geometry is realized in our macroscopic pilot-wave experiment by considering a walker confined to an annulus that encircles a shielded vortex. 
Crucially, the droplet’s trajectory lies entirely outside the vortex core, where the mean surface flow is negligible; thus, analogous to the quantum Aharonov--Bohm effect, no local forces act on the particle. 
Instead, the vortex modifies the droplet’s extended pilot-wave field, which spans the annulus and overlaps the vortex region. Because the droplet’s propulsion is set by the local slope of this pilot-wave at the point of impact, a vortex-induced phase shift in the wave translates into a systematic shift of the droplet’s effective (canonical) momentum, even in the absence of any direct hydrodynamic forcing along the trajectory.
High-speed tracking and delay-embedding tomography yield an empirical $W(x,p)$ that exhibits the same rigid momentum shift $\Delta p$ as in Eq.~\eqref{eq:Deltap}, providing a direct, single-particle phase-space signature of the AB effect (see Supplementary Information for reconstruction details).

\begin{figure}
    \centering
    \scalebox{0.65}{\includegraphics{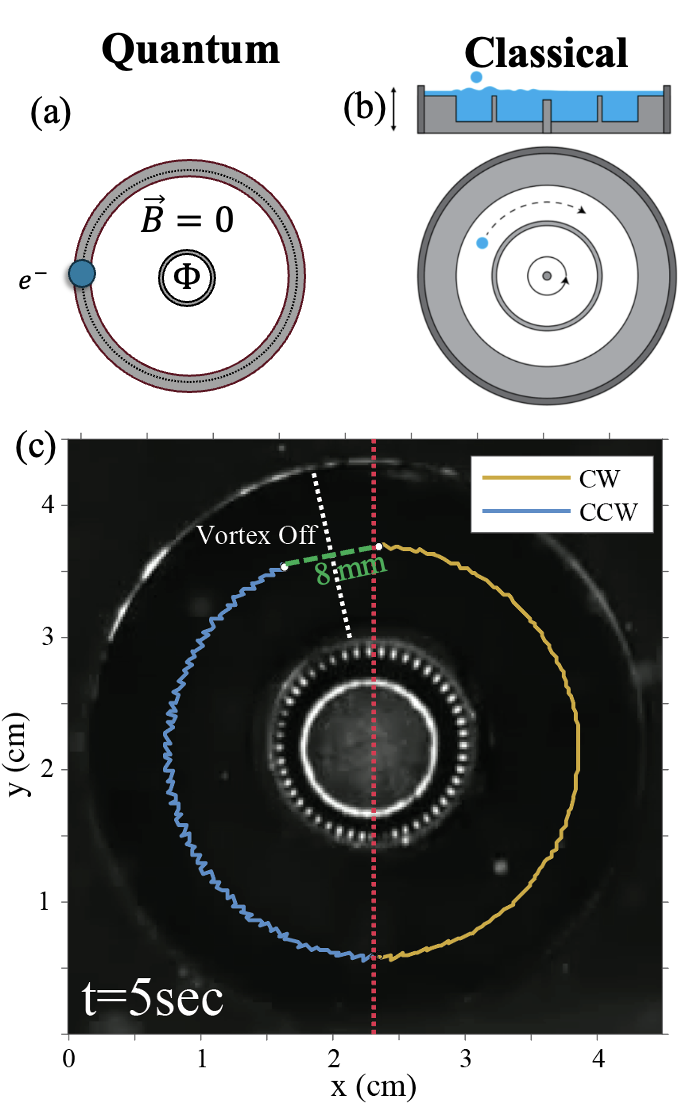}}
     \caption{(a) Schematic of the Aharonov--Bohm (AB) effect for an electron constrained to a ring threaded by magnetic flux~$\Phi$. 
(b) Hydrodynamic analogue: a walking droplet (``walker'') travels along an annular cavity while a weak, centrally pinned vortex generates a synthetic magnetic flux. 
(c) Experimental images showing two representative trajectories of walkers circulating in the annulus while the central vortex rotates clockwise.}
\label{fig:1}
\end{figure}

\begin{table*}[t]
\centering
\caption{Correspondence between the quantum Aharonov--Bohm effect and its hydrodynamic pilot-wave analogue.}
\label{tab:AB-comparison}
\setlength{\tabcolsep}{4pt} 
\renewcommand{\arraystretch}{1.05} 
\setlength{\tabcolsep}{4pt}
\resizebox{0.95\textwidth}{!}{%
\begin{tabular}{lcc}
\toprule
\textbf{Property} &
\textbf{Quantum Aharonov--Bohm effect} &
\textbf{Hydrodynamic analogue (this work)} \\
\midrule
Topology &
Ring enclosing magnetic flux &
Annulus enclosing central vortex \\

Local field on trajectory &
$\mathbf{B}=0$ &
Mean surface flow $\approx 0$ \\

Gauge potential &
Electromagnetic vector potential $\mathbf{A}$ &
Synthetic azimuthal gauge potential \\

Flux / control parameter &
Magnetic flux $\Phi=\oint \mathbf{A}\cdot d\boldsymbol{\ell}$ &
Vortex control parameter $\Gamma_v$ \\

Phase (holonomy) &
$\Delta\varphi = (q/\hbar)\,\Phi$ &
$\Delta\varphi_{\rm eff} \propto \Gamma_v$ \\

Canonical momentum shift &
$\Delta p = q\Phi/(2\pi R)$ &
$\Delta p_{\rm eff} \propto \Gamma_v$ \\

Phase-space signature &
Rigid translation of Wigner function &
Rigid translation of empirical $W(x,p)$ \\
\bottomrule
\end{tabular}}
\end{table*}

\paragraph*{Experimental settings.}
Experiments were conducted in a shallow annular cavity machined from aluminum and filled with 20~cSt silicone oil 
($\rho=0.95$~g\,mL$^{-1}$, $\sigma=20.6$~mN\,m$^{-1}$, $\nu=20$~cSt). 
The annulus had an outer diameter of 85~mm and an inner diameter of 45~mm. 
Here $\lambda_F$ denotes the Faraday wavelength of the parametrically excited surface waves at the forcing frequency.
Expressed in these units, $R_{\mathrm{out}}\simeq4.3\,\lambda_F$ and $R_{\mathrm{in}}\simeq2.2\,\lambda_F$, and the bath depth was $h=10.0\pm0.2$~mm. 
A circular disk of diameter 10~mm was mounted concentrically at the center and submerged 
$2.0\pm0.4$~mm below the unperturbed liquid surface to generate a weak, localized vortex flow. 
The bath was driven sinusoidally at frequency $f=80$~Hz with a peak acceleration $\gamma=3.95\,g$, 
corresponding to a reduced acceleration $\Gamma=\gamma/\gamma_F=0.96$ (see Supplementary Information for the memory-parameter definition).
The Faraday wavelength was measured to be $\lambda_F=4.6\pm0.1$~mm, in agreement with previous reports 
for 20~cSt silicone oil at this frequency.

Monodisperse droplets of radius $R_0 = 0.45 \pm 0.03$~mm 
(mass $m \simeq 3.2$~mg) were produced by needle retraction and allowed to reach steady walking before data acquisition. 
All measurements were conducted under a transparent lid to suppress air currents and ensure thermal stability.

\paragraph*{Operating conditions.}
To isolate the influence of the synthetic flux on the walker’s dynamics,
we examined four distinct regimes combining the presence or absence of the central vortex
with the direction of the walker’s orbital motion in the annulus.
The vortex was generated by a submerged rotating disk driven clockwise at a fixed setting
($f_v = 54$~Hz, $\omega_v \simeq 339$~rad\,s$^{-1}$).

The vortex strength was controlled experimentally by the signed drive voltage $V$
applied to the disk motor.
For convenience, we introduce a dimensionless control parameter $\Gamma_v$,
defined as the motor drive voltage expressed in normalized units,
$\Gamma_v \propto V$,
with $\Gamma_v=0$ corresponding to the vortex-off state and $\Gamma_v<0$ denoting clockwise rotation.
Over the operating range used here, the disk rotation rate $\Omega$ depends linearly on $V$,
so that $\Gamma_v$ is also linearly proportional to $\Omega$.
Throughout the vortex-on data shown in this work, the calibrated value is $\Gamma_v=-0.32$.

The walker’s orbital motion is characterized by the normalized angular velocity
\(\hat{\Omega}=2\Omega/\omega_F\), where \(\Omega\) is the measured orbital rotation rate and
\(\omega_F\) is the Faraday angular frequency.
Positive \(\hat{\Omega}\) corresponds to clockwise motion, and the magnitude
\(|\hat{\Omega}_0|\) was held fixed across all experiments.
We consider four operating regimes obtained by combining the presence or absence of the central vortex
(\(\Gamma_v=0\) or \(-0.32\)) with co-rotating (\(\hat{\Omega}=+\hat{\Omega}_0\)) and counter-rotating (\(\hat{\Omega}=-\hat{\Omega}_0\)) orbital motion.

\begin{figure*}[ht]
    \centering
    \scalebox{0.35}{\includegraphics{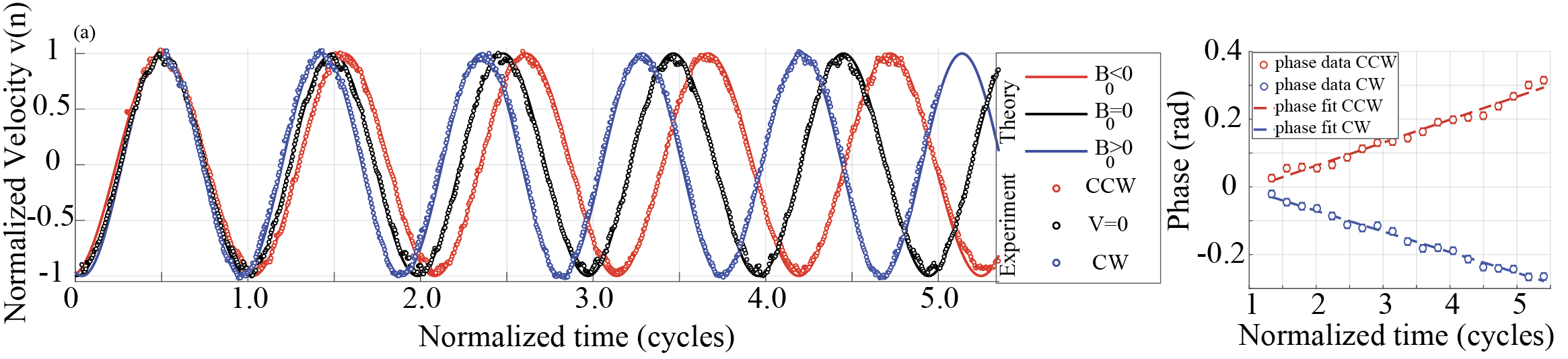}}
     \caption{(a) Overlay of the normalized longitudinal velocity $v_x(t)$ for three regimes: clockwise (CW), vortex-off baseline (VOFF), and counterclockwise (CCW). Experimental data (markers) are normalized to unit amplitude and aligned in cycles, and are compared directly with the Aharonov--Bohm theoretical reference $\langle v_x(t)\rangle=v_0\cos(\omega t)$, with $v_0=\frac{\hbar n}{mR}-\frac{q\Phi}{2\pi m R}$ and $\omega=v_0/R$ (solid curves). (b) Relative phase accumulation versus period (cycles, $\tau$). Instantaneous phases are extracted from the normalized $v_x$ traces; differences are formed (CW--VOFF in red; VOFF--CCW in blue) and binned uniformly (20 bins) with SEM error bars. Dashed lines show linear fits whose slopes give the net phase-accumulation rates. For the co-rotating case (red), the fit yields $\Delta\varphi_{\mathrm{CW}}(\tau)=0.068\,\tau-0.073$, corresponding to a positive, flux-induced phase drift per cycle. For the counter-rotating case (blue), the fit yields $\Delta\varphi_{\mathrm{CCW}}(\tau)=-0.062\,\tau+0.052$, demonstrating an opposite-sign phase accumulation of comparable magnitude. The vortex-off baseline remains consistent with zero phase drift, confirming that the observed phase accumulation arises from the synthetic gauge flux.}
    \label{fig:2-velandphase}
\end{figure*}

\paragraph*{Velocity diagnostics.}

Figure~\ref{fig:2-velandphase} summarizes the kinematic signature of the AB analog in our annular geometry. We focus on the longitudinal velocity component $v_x(t)$ rather than the speed $|v|$ because $v_x(t)$ retains the sign and phase information of the orbital motion, whereas $|v|$ removes this information and obscures any accumulated phase shift. Panel~(a) overlays the normalized longitudinal velocity traces $v_x(t)$ for clockwise (CW), vortex-off (VOFF), and counterclockwise (CCW) runs and compares them with the corresponding Aharonov--Bohm reference signal, whose functional form is given in Eq.~\eqref{eq:vx_expectation}.

To facilitate a direct comparison between reference and experiment, both signals are expressed in terms of a dimensionless cycle variable $\tau$: $\tau_{\mathrm{th}}=(t-t_0)/T_0+1$ for the reference and $\tau_{\mathrm{exp}}=(t-t_0)/T_{\mathrm{exp}}+1$ for the experiment, with $T_0=5.5\times10^{-8}\,\mathrm{s}$ and $T_{\mathrm{exp}}=10\,\mathrm{s}$, respectively; amplitudes are scaled to unit maximum. Panel~(b) quantifies the relative phase accumulation by extracting the instantaneous phase of each $v_x(t)$ trace and comparing phase differences between regimes (see Supplementary Information for details). Linear trends (dashed) yield the phase-accumulation rates, which reverse sign between co-rotating and counter-rotating motion, while the VOFF baseline remains statistically flat. This phase drift occurs despite the vortex-induced circulation being zero along the particle trajectory and arises from the vortex-induced modification of the droplet’s pilot-wave field that guides its motion.

\paragraph*{Phase-space formulation.}


While the Aharonov--Bohm effect may be discussed in either position or momentum space, these complementary representations capture only partial aspects of the underlying gauge physics. Position-space interference reveals flux-dependent phase shifts, while momentum-space observables encode corresponding shifts of dynamical quantities. A phase-space formulation provides a more complete description by making explicit how the gauge potential enters as a rigid translation of the canonical momentum.
In a single-branch preparation (clockwise population only), the AB effect is most naturally expressed in terms of the Wigner function, a quasiprobability distribution in phase space that retains simultaneous information about position and momentum~\cite{Wigner1932}. For a Gaussian packet of widths $(\sigma_{s},\sigma_{p})$ centered at $(s_{0},p_{0})$, the Aharonov--Bohm flux produces a uniform translation along the momentum axis,
\begin{equation}
W_{\text{AB}}(s,p)=\frac{1}{2\pi\sigma_{s}\sigma_{p}}\,
\exp\!\left[-\frac{(s-s_{0})^{2}}{2\sigma_{s}^{2}}-\frac{\bigl(p-(p_{0}+\Delta p)\bigr)^{2}}{2\sigma_{p}^{2}}\right],
\label{eq:WignerShift}
\end{equation}
with $\Delta p$ given by Eq.~\eqref{eq:Deltap}. The Aharonov--Bohm phase thus appears in phase space as a flux-dependent, rigid translation of $W$ along the momentum direction, while the spatial profile remains unchanged~\cite{Cembranos2024,schleich2011quantum}.

\paragraph*{Phase-space reconstruction.}

\begin{figure}[ht]
    \centering
    \scalebox{0.4}{\includegraphics{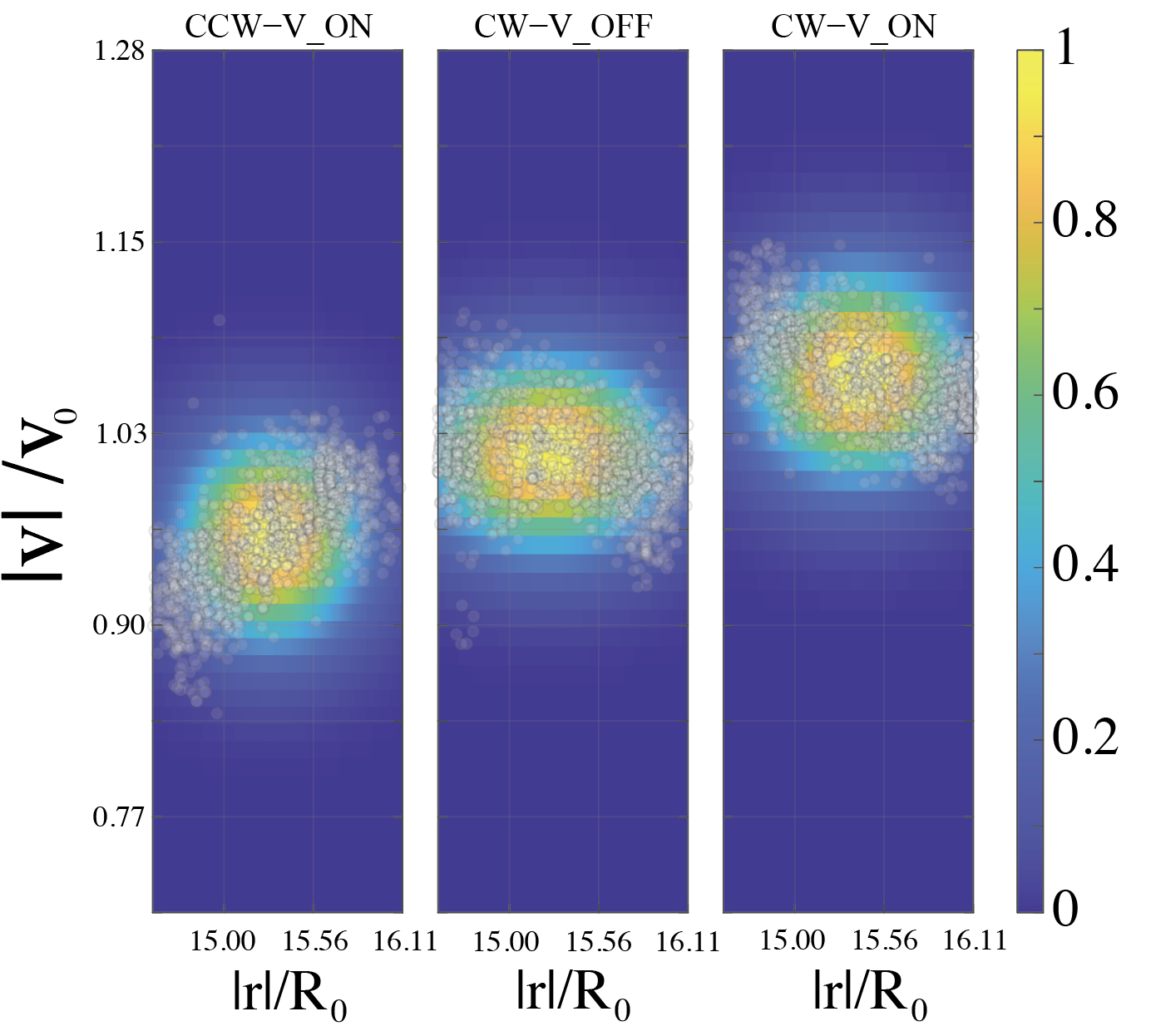}}
     \caption{Wigner-like phase-space distributions $W(x,p)$ reconstructed from droplet trajectories using delay-embedding tomography with lag $\tau = T_F/2$. 
Panels~(a)–(c) correspond, respectively, to the counter-rotating state (lowest momentum), the vortex-off baseline ($\Gamma_v = 0$), and the co-rotating state (highest momentum). 
The baseline distribution exhibits an approximately isotropic Gaussian envelope centered at $(\bar{x},\bar{p})$, consistent with the theoretical form of a single-branch Gaussian packet. 
Activating the vortex produces a rigid translation of the distribution along the momentum axis by $\Delta p/p_0 \approx \pm 0.05$, with no measurable change in width or covariance. This flux-dependent displacement constitutes a direct phase-space analogue of the Aharonov--Bohm momentum shift and quantitatively matches the phase offset inferred from the velocity-based analysis in Fig.~\ref{fig:2-velandphase}.}
    \label{fig3:Wigner_2D}
\end{figure}

Using high-speed trajectory data and a delay-embedding reconstruction with lag $\tau = T_F/2$, 
we obtained Wigner-like distributions $W(x,p)$ for $600$ impacts per regime, extracted from long, steady-state droplet trajectories shown in Fig.~\ref{fig3:Wigner_2D}(a–c). The Wigner-like phase-space distributions are reconstructed from a single, long droplet trajectory using a delay-embedding procedure, rather than from an ensemble of independent realizations. Panel~(a) corresponds to the lowest-momentum state (counterclockwise orbit, vortex on), 
panel~(b) to the zero-flux baseline ($\Gamma_v = 0$), 
and panel~(c) to the highest-momentum state (clockwise orbit, vortex on). 
The baseline distribution is centered at $(\bar{x},\bar{p})$ and exhibits an approximately isotropic Gaussian envelope 
consistent with the theoretical form of Eq.~\eqref{eq:WignerShift}.
Introducing the vortex produces a rigid translation of the distribution along the momentum axis 
by $\Delta p \approx \pm 0.05\,p_0$, without measurable change in width or covariance, 
\textit{demonstrating that the synthetic flux acts as a pure gauge potential.} Within experimental resolution, the effect of the vortex is a rigid translation of the distribution along the momentum axis, with no detectable broadening or deformation. This shift corresponds to an inferred effective flux $\Phi_{\mathrm{eff}}/\Phi_0 = 0.05 \pm 0.01$, consistent with the phase offset deduced from the velocity comparison in Fig.~\ref{fig:2-velandphase}.

\paragraph*{Flux-dependence of the momentum shift.}
A central prediction of the Aharonov--Bohm effect is the linear dependence of the canonical-momentum shift on the enclosed flux. To test the corresponding scaling in our hydrodynamic analogue, we varied the rotation rate of the submerged disk, and thus the strength of the synthetic vortex flux, over a broad range and measured the resulting orbital speeds of the walker. For each imposed disk frequency, long trajectories were recorded and the mean speed $|v|$ was extracted in both co-rotating (CW) and counter-rotating (CCW) directions. The vortex strength was calibrated independently (see Supplementary Information).

\begin{figure}
    \centering
    \scalebox{0.4}{\includegraphics{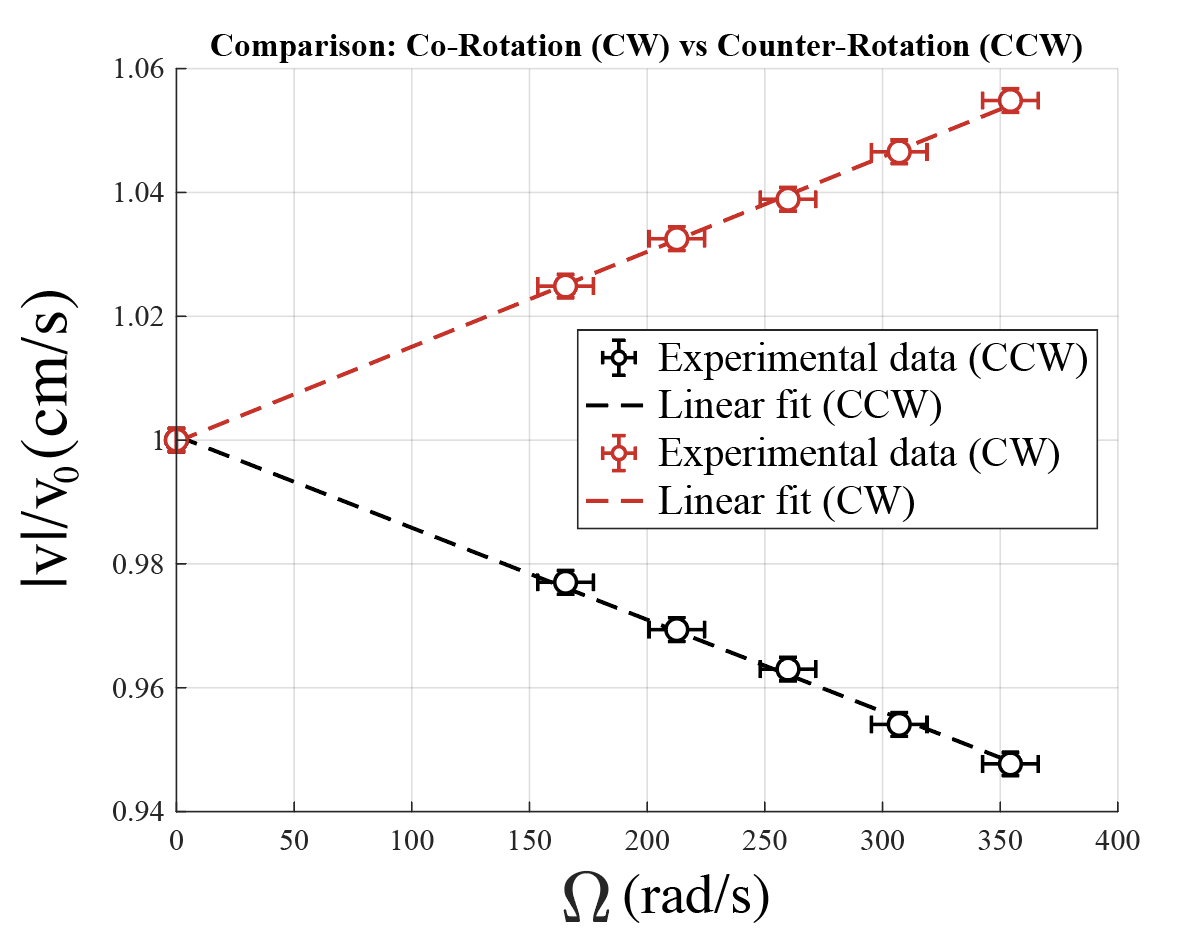}}
     \caption{Dependence of mean orbital speed $|v|$ on vortex rotation rate $\Omega$
    for co-rotating (red) and counter-rotating (black) motion.
    Error bars denote SEM. Linear fits (dashed) show nearly antisymmetric slopes,
    indicating that the synthetic gauge flux is proportional to $\Omega$}
    \label{fig:4}
\end{figure}

Figure~\ref{fig:4} shows the dependence of the orbital speed on $\Omega$ for both circulation directions. 
In the CW configuration, the mean speed increases approximately linearly with vortex strength, while in the CCW configuration it decreases by a similar amount. Linear fits yield
\begin{equation}
v_{\rm CW/CCW}(\Omega)
= v_0 \pm (1.20,\;1.16)\times10^{-4}\,\Omega,
\label{eq:vOmega}
\end{equation}

The fits in Eq.~\eqref{eq:vOmega} quantify how the imposed vortex circulation biases the orbital motion. 
The slopes give the speed (and hence canonical-momentum) shift per unit vortex rotation rate $\Omega$, while the intercept $v_0$ defines the zero-flux orbital speed. 
The near-antisymmetry of the CW and CCW slopes indicates a uniform, tunable gauge-induced bias rather than a local forcing or nonlinear flow effect. 
The two branches intersect at $\Omega=0$ within uncertainty, indicating no intrinsic directional bias when the vortex is off. 
Here $v_0 \equiv 0.784\pm0.002~\mathrm{cm\,s^{-1}}$ is the measured zero-flux speed.

Since momentum is proportional to speed for the nearly constant droplet mass ($p \propto v$), our results imply a flux-dependent momentum shift $\Delta p/p_{0}=\Delta v/v_{0}\approx0.06$ at the highest vortex rotation rates considered ($\Omega\simeq350~\mathrm{rad/s}$). In the quantum reference, the characteristic velocity scale $v_0$ is set by the difference between the mechanical momentum $\hbar n/(mR)$ and the gauge contribution $q\Phi/(2\pi mR)$ [Eq.~\eqref{eq:flux}], so that variations in the enclosed flux produce a linear shift of the velocity. In the hydrodynamic analogue, the vortex rotation rate $\Omega$ plays the role of a tunable control parameter for an effective flux $\Phi_{\mathrm{eff}}$, leading to an analogous linear bias of $v_0$, the droplet’s orbital speed. The quantitative consistency across independent diagnostics therefore demonstrates that the submerged vortex generates a genuine AB-type gauge phase whose strength is linearly tunable.

\section{Discussion and Conclusion}

Our results demonstrate that a walking droplet confined to an annular cavity acquires a flux-dependent dynamical bias that is fully analogous to the Aharonov--Bohm effect. Three independent diagnostics are reported: (i) the phase accumulation visible in the longitudinal velocity traces, (ii) the rigid momentum-space translation extracted from the empirical Wigner distributions, and (iii) the linear divergence of co- and counter-rotating speeds under controlled variation of the vortex strength. All are consistent with a single underlying mechanism: a uniform azimuthal phase shift induced by the vortex-imposed modification of the pilot-wave field.
Crucially, the droplet trajectory lies entirely outside the vortex core, where direct hydrodynamic forcing is negligible, and control measurements show no detectable drift of stationary bouncers (see Supplementary Information).
The observed momentum shift is therefore not a consequence of local forcing but rather of a global, path-dependent phase imprinted on the droplet dynamics through its pilot wave. This interpretation precisely mirrors the AB effect, where the phase shift arises from a gauge potential despite the absence of magnetic field along the particle path \cite{Peshkin2005}.
Beyond verifying the hydrodynamic Aharonov–Bohm analogy, the phase-space perspective provides a more refined insight: the entire Wigner-like distribution undergoes a rigid translation without detectable broadening or deformation.
The quantitative agreement between the Wigner shift, the velocity-based diagnostics, and the flux-sweep slopes shows that this table-top classical system reproduces not only qualitative AB-like features, but the correct gauge-induced phase-space response expected for a flux-threaded ring.
These results establish walking droplets as a powerful platform for studying synthetic gauge fields and a path-dependent (geometric) phase in the sense of the Aharonov–Bohm effect, in a regime where full, trajectory-resolved phase-space reconstruction is experimentally accessible. Future work could extend this approach to probe decoherence and dephasing.

\section*{Acknowledgments}
JB acknowledges the support of the Office of Naval Research (ONR) through grant N00014-24-1-2232
and the National Science Foundation (NSF) through grant CMMI-2154151.
G.G.R.\ acknowledges additional support from the MIT School of Science Research Innovation Seed Fund.

$\,$

$\,$

\end{document}